\documentclass[conference]{IEEEtran}
\usepackage{graphicx,url} 

\usepackage[top=1in, bottom=1in, left=1in, right=1in]{geometry}

\usepackage[linesnumbered,ruled,vlined]{algorithm2e}
\SetKwInput{KwInput}{Input}                
\SetKwInput{KwOutput}{Output}              
\usepackage{stfloats,xcolor}
\usepackage{enumitem,amsmath}
\newtheorem{property}{Property}
\newtheorem{definition}{Definition}
\newcommand{\ignore}[1]{}
\newcommand{\mcf}{\ensuremath{\sf mcf}}
\newcommand{\mcfg}{\ensuremath{\sf mcfg}}
\newcommand{\mdf}{\ensuremath{\sf mdf}}
\newcommand{\mdfg}{\ensuremath{\sf mdfg}}
\newcommand{\mission}{\ensuremath{\sf mission}}

\newcommand{\general}{\ensuremath{\sf general}}
\newcommand{\A}{\ensuremath{{\cal A}}}

\begin{document}

\title{Towards Principled Risk Scores for Space Cyber Risk Management}

\author{\IEEEauthorblockN{Ekzhin Ear}
\IEEEauthorblockA{\textit{Department of Computer Science} \\
\textit{Uni. of Colorado Colorado Springs}\\
Colorado Springs, United States \\
eear@uccs.edu}
\and
\IEEEauthorblockN{Brandon Bailey}
\IEEEauthorblockA{\textit{Cyber Assessments and Research} \\
\textit{The Aerospace Corporation}\\
El Segundo, CA, United States \\
brandon.bailey@aero.org}
\and
\IEEEauthorblockN{Shouhuai Xu}
\IEEEauthorblockA{\textit{Department of Computer Science} \\
\textit{Uni. of Colorado Colorado Springs}\\
Colorado Springs, United States \\
sxu@uccs.edu}
}

\date{}


\maketitle

\begin{abstract}
Space is an emerging domain critical to humankind. Correspondingly, space cybersecurity is an emerging field with much research to be done. To help space cybersecurity practitioners better manage cyber risks, The Aerospace Corporation proposed Notional Risk Scores (NRS) within their Space Attack Research and Tactic Analysis (SPARTA) framework, which can be applied to quantify the cyber risks associated with space infrastructures and systems. While intended for adoption by practitioners, NRS has not been analyzed with real-world scenarios, putting its effectiveness into question. In this paper we analyze NRS via a real-world cyber attack scenario against a satellite, and characterize the strengths, weaknesses, and applicability of NRS. The characterization prompts us to propose a set of desired properties to guide the design of future NRS. As a first step along this direction, we further propose a formalism to serve as a baseline for designing future NRS with those desired properties.
\end{abstract}


\section{Introduction}

Real-world cyber attacks against space infrastructures and systems have been reported for over four decades \cite{ear2023characterizing, pavur2022building, falco2021security}. However, space cybersecurity practitioners still lack tools to effectively understand and manage cyber risks associated with space infrastructures and systems, or {\em space cyber risks} in short. In a major effort to support space cybersecurity practitioners, The Aerospace Corporation developed and incorporated space cyber {\em Notional Risk Scores} (NRS) \cite{sparta_notional} into their Space Attack Research and Tactic Analysis (SPARTA)
framework {\cite{sparta_homepage}, associating a notional evaluation of cyber risks to attack techniques.
NRS and SPARTA are founded on a wealth of industry research and precedent, including traditional enterprise Information Technology (IT) cybersecurity (e.g., 
security controls \cite{NIST800-53r5,NIST800-53A,NIST800-53B}). 

The intention of NRS is to provide practitioners with a starting point for space cyber risk management, from which they can apply specific details (e.g., a reference architecture) to tailor NRS to evaluate their particular cyber risks. However, the precise methodology for practitioners to apply NRS is scant. Further, NRS has not been analyzed with real-world scenarios, putting its effectiveness in question. This motivates us to characterize the strengths, weaknesses, and applicability of NRS. 


\noindent{\bf Our Contributions}.
In this paper we make three  contributions.
First, we present an algorithmic  description 
of applying NRS to conduct space cyber risk management. 
This is both important and useful because the current description of NRS \cite{sparta_notional} does not thoroughly explain the required details for practitioners to correctly understand, adopt, and apply NRS. 
The algorithmic description paves the way for us to objectively characterize the strengths, weaknesses, and applicability of NRS
via a case study on a real-world cyber attack against a satellite and its associated systems. 

Second, in order to guide the development of future versions of NRS, we propose a comprehensive set of desired properties that should be satisfied by future NRS. This is important to avoid ad hoc development 
of future NRS because these properties, or their future refinements, can be used to assess the competency of any NRS, including the current one and its future  developments.

Third, as a first step towards developing a desired NRS, we propose a formalism which is centered at three key concepts: (i) {\em mission}, which specifies the tasks supported by a space system in question;
(ii) {\em mission control flow}, which specifies the chains of commands from one unit of a space infrastructure/system to another that are needed to accomplish a mission; (iii) {\em mission data flow}, which specifies how the data items are transmitted between, and processed by, space infrastructure/system units. 
The formalism serves as a starting point 
for developing future NRS with the desired properties and automated space cyber risk management tools.





\noindent{\bf Related Work}.
Since this study focuses on NRS and space cyber risk management, we review related prior studies in a broader vs. the space context.
In a broader context, there are two major cyber risk management frameworks:
(i) the National Institute of Standards and Technology (NIST) Cybersecurity Framework (CSF) \cite{NISTCSF}, which provides an overarching structure with five cybersecurity functions;
and, (ii) 
NIST Risk Management Framework (RMF) \cite{joint2018nist800-37}, which implements NIST CSF and contains a system lifecycle approach for security
that use security controls
\cite{NIST800-53r5,NIST800-53A,NIST800-53B,joint2012nist800-30}. 
These endeavors are leveraged by NRS to support the adaption of IT security controls to space-related countermeasures, meaning that these IT network security controls
could be applied to secure
space infrastructures and systems.


Specific to space, NIST \cite{scholl2021introduction} provides guidance for
creating CSF profiles for space operations, and applies it for
(i) positioning, navigation, and timing (PNT) satellite services \cite{bartock2021foundational}, (ii) satellite command and control systems in the ground segment \cite{lightman2022satellite}, and (iii) hybrid satellite networks 
\cite{mccarthy2023cybersecurity}. 
Moreover, NASA \cite{NASA-STD-1006A} establishes perhaps the first set of high-level security controls to maintain command authority, protect PNT, and assure interference reporting. These controls can be applied to reduce space cyber risks \cite{jones2022risk,NASA-SS-BPG, bailey2021cybersecurity},
and are leveraged by NRS.

Related academic research on space cyber risks includes \cite{knez2016lessons}, which reports lessons learned from their process of applying RMF
to a notional space mission, while discussing {\em some} desired properties of cyber risk assessments.
In contrast, we propose a {\em comprehensive} set of desired properties for NRS.
Ear et al. \cite{ear2023characterizing} study real-world cyber attacks against space infrastructures and systems while proposing a framework that includes a comprehensive space system model and two metrics: {\em attack consequence} and {\em attack sophistication}.
The formalism proposed in the present study can be incorporated into their framework as a building-block.

\noindent{\bf Paper Organization}.
Section \ref{sec:notional-risk-score-v1} reviews NRS. Section \ref{sec:characterizing-v1.0} leverages a case study to characterize its strengths, weaknesses, and applicability. Section \ref{sec:desired-notional-risk-score} presents desired properties of NRS.
Section \ref{sec:notional-risk-score-v2} presents a formalism to support the development of future NRS with the desired properties.
Section \ref{sec:conclusion} concludes the paper.

\section{A Review of NRS}
\label{sec:notional-risk-score-v1}

\noindent{\bf Background}. The MITRE 
ATT\&CK framework \cite{sparta_matrix} is geared towards terrestrial networks. To establish a similar framework for space infrastructures and systems, The Aerospace Corporation
proposed the SPARTA framework \cite{sparta_homepage}. The two frameworks are complementary to each other in terms of their application settings. 
In August 2023, The Aerospace Corporation further proposed NRS \cite{sparta_notional}
in an update to its SPARTA framework. 
NRS is based on \cite{bailey2020establishing,bailey2021cybersecurity}, which leverages NIST SP 800-53 \cite{joint2012nist800-53} to 
create a 
threat-based approach to space cyber risk management.
SPARTA and
NRS have been adopted by some space cyber practitioners.

\subsection{Space System Model and Terminology}

\noindent{\bf Space infrastructure system model}. 
To avoid ambiguity, we need a clearly-defined space infrastructure system model, which is currently not offered by SPARTA 
nor by NRS.
This prompts us to adopt the space system model recently presented by researchers in \cite{ear2023characterizing} because it is the most comprehensive space infrastructures and systems model we find from the literature. The model is depicted in Figure~\ref{fig:systemModel}.
At a high level, space infrastructures contain four segments, namely {\em space}, {\em ground}, {\em user}, and {\em link}; each segment contains one or multiple components (e.g., the space segment consists of two components: {\em bus system} and {\em payload}); each component contains multiple modules (e.g., the bus system component contains six modules). We refer to \cite{ear2023characterizing} for detailed descriptions of segments, components, and modules.

\vspace{-1em}
\begin{figure}[!htbp]
\centering{\includegraphics[width=\columnwidth]{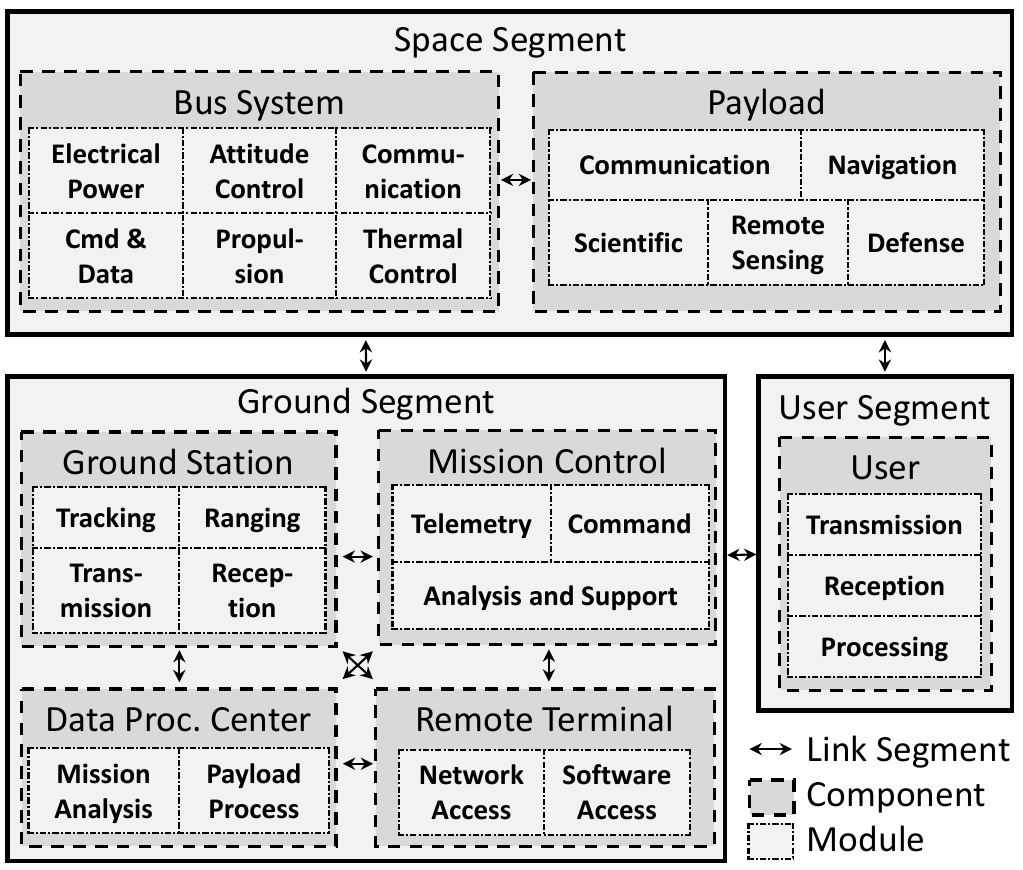}}
\vspace{-2em}
\caption{Space infrastructure system model (adapted from \cite{ear2023characterizing}).}
\label{fig:systemModel}
\vspace{-.6em}
\end{figure}

\noindent{\bf Terminology}. 
For describing cyber attacks against, and defenses for, space infrastructures and systems with respect to the system model described above, we adopt the following terminology used by SPARTA and NRS \cite{sparta_matrix, sparta_defcon, sparta_countermeasures}.
{\em Tactics} are cyber threat actors' goals or objectives.
{\em Techniques} are the actions cyber threat actors take to accomplish their tactics.
{\em Sub-techniques} are 
more specific instances of a parent technique.
{\em Procedures} are the step-by-step implementations of (sub-)techniques. 
{\em Attack chains} are ordered sets of SPARTA techniques employed by attackers to accomplish their attack objectives.
{\em Countermeasures} are protective measures, which are instantiated as {\em security controls}, namely
safeguards for protecting systems and data. 
NIST \cite{NIST800-53r5} provides 20 families of 
security controls.

To characterize NRS, we introduce the following terms. {\em Attack flows} are ordered sets of space system components/modules that adversaries must compromise in order to achieve their objectives. 
Note that {\em attack flows} are related to, but different from, {\em attack chains}. {\em Mission control flows} are ordered sets of space infrastructure units (including links) required to send commands that accomplish a mission. Note that the word ``control'' in {\em mission control flow} has nothing to do with the same word in ``security control'' because the former deals with functions and the later deals with security.
{\em Mission data flows} are ordered sets of space infrastructure units and links required to send data items according to a mission.


\subsection{Overview of the Space Cyber NRS}

The SPARTA framework  provides 
technical information about how cyber attackers may compromise spacecraft
(e.g., satellites) and 
helps space cyber defenders understand space cybersecurity through the lens of tactics, techniques, and procedures (TTPs). It provides nine tactics; each tactic contains between 5 to 18 techniques; some techniques contain between 1 to 9 sub-techniques; and, each (sub-)technique is mapped to countermeasures and
NIST SP 800-53 
security controls \cite{NIST800-53r5}.

\ignore{
{\color{green}
SPARTA has the following target audiences \cite{sparta_leveraging}:
(i) space system developers,
who can specify and engineer 
countermeasures against SPARTA techniques into their systems' lifecycle; 
(ii) space defensive cyber operations (DCO) analysts, who can construct and assess defense mechanisms based on the applicable TTPs found in SPARTA;
(iii) threat intelligence analysts, who can produce raw, polished, and aggregated reporting by referencing SPARTA;
(iv) assessment engineers, who can 
use SPARTA techniques to identify viable cyber attack chains; and,
(v) educators, trainers, and researchers, who can use SPARTA to convey space cyber knowledge.
}
}



\ignore{
The developers of SPARTA introduced the 
Notional Risk Scores (NRS) as an integration into the SPARTA framework
as a starting point for all
of the five target audiences mentioned above to quantitatively assess their risk in terms of the impact of adversaries' employment of techniques described by SPARTA  \cite{sparta_notional}. 
}

At a high level, NRS provides, for each system criticality, base risk scores for each SPARTA technique, determined by 
(i) technique likelihood of successful attack and (ii) technique impact incurred, via a risk matrix.
In addition, NRS specifies countermeasures and security controls for mitigating specific SPARTA techniques.

\subsubsection{System Criticality}
There are three categories in system criticality: 
{\em high}, associated with systems related to critical functions,
military purposes, and intelligence activities; {\em medium}, associated with civil, science, weather, and commercial systems; and, {\em low}, associated with academic and research systems \cite{sparta_riskassessment}. 
Aerospace Corporation experts assign a base risk score for each SPARTA technique at each system criticality level.
Practitioners subjectively determine the criticality level of each space system unit impacted by a SPARTA technique to attain the SPARTA technique's base risk score. 

\subsubsection{SPARTA Technique Likelihood}
The evaluation of technique likelihood includes three aspects: (i) adversary motivation, influenced by the system criticality with the assumption that adversaries are more motivated to attack {\em high} criticality rather than {\em low} criticality systems; (ii) exploitation difficulty, based on technique complexity; and, (iii) adversary capabilities, according to the following seven tiers, in increasing order: script kiddies, hackers for hire, small hacker teams, insider threats, large well-organized teams, highly capable state actors, and most capable state actors \cite{bailey2021cybersecurity}.
Subjective analysis on these three aspects 
provides the overall likelihood score which results in a range $\{1,\ldots,5\}$.


\subsubsection{SPARTA Technique Impact}
The impact of a technique against a space system refers to the consequences, effects, or outcomes resulting from the successful execution of the technique. 
Subjective analysis considers 
wide-ranging impact that may include mission disruption, data integrity, loss of control or availability, financial consequences, safety, or even national security implications \cite{sparta_riskassessment}. Impact is also defined in a range $\{1,\ldots,5\}$.

\subsubsection{Risk Matrix Representation (Risk Scores)}
This is a 5$\times$5 risk matrix 
representation of the notional risk scores of the SPARTA techniques evaluated \cite{NASA-S3001}. The matrix 
provides a risk score with respect to an assessed {\em impact} score from 1 to 5 (the $x$-axis) and a {\em likelihood} score from 1 to 5 (the $y$-axis); the risk scores are shown in the respective cells of the matrix and reflect the joint effect of impact and likelihood, according to the 5$\times$5 matrix defined in \cite{NASA-S3001}.
Risk scores range from 1 to 25, but are {\em not} the product of likelihood and impact.
Risk scores ranging from 1 to 10 are considered {\em low}, 11 to 19 considered {\em medium}, and 20 to 25 considered {\em high}.

\subsubsection{Countermeasures and Security Controls} 
These are integrated into SPARTA and can be employed to thwart SPARTA techniques.
Each countermeasure contains space-specific protection measures, which can be instantiated by 
NIST SP 800-53 security controls to achieve the goal of the countermeasure.
For example, SPARTA technique {\em Memory Compromise (PER-0001)} is mapped to eight countermeasures, meaning that an attack exploiting the technique can be thwarted by one or a combination of multiple countermeasures. For instance, the countermeasure with identifier {\em CM0021} specifies that the digital signature of flight software is verified prior to installation, while noting that {\em CM0021} is mapped
to 19 security controls (e.g., security control {\em CM-11} specifies that organizational policies governing the installation of software must be established, enforced, and monitored).

\subsection{Algorithm for Using NRS}


\begin{algorithm}[!ht]
    \label{alg:NRS}
\caption{Using NRS for risk management}
    \KwInput{SPARTA matrix of techniques; {\em subjective} NRS base risk score $R$, which is a set indexed by technique with element $R_A$ being the basic risk score of technique $A$; a set indexed by SPARTA technique with element $C_A$ being a set of countermeasures to technique $A$; a set indexed by countermeasure, with element $S_{c_A}$ being a set of secure controls fulfilling countermeasure $c_A\in C_A$; specific environment/conditions of the space system; tolerable risk threshold $\tau\in\{`low', `medium', `high'\}$}
    \KwOutput{a set of security controls that must be employed to mitigate intolerable risks}


    {\em subjectively} determine the set $\A$ of SPARTA techniques applicable to the space system

    \For{{\em each technique} $A\in \A$} 
    {
    identify its base risk score $R_A$ according to $R$
    
    {\em subjectively} tailor the base risk score to $r_A=(impact,likelihood)$ to reflect the specific environment/conditions

        
     
    populate $A$ on the 5$\times$5 risk matrix
    
    
        \If{{\em the risk incurred by} $A>\tau$}
        {

            {\em subjectively} select a set of countermeasures to mitigate $A$, denoted by $\{c_A\}\subseteq C_A$

            {\em subjectively} select a set of security controls to fulfill the countermeasure, denoted by $S_{\{c_A\}}$

        }
    }
    
\Return{$\bigcup_{A\in \A} S_{\{c_A\}}$}   

\end{algorithm}

Algorithm \ref{alg:NRS} shows how to use NRS to quantify space cyber risk and identify mitigation, which was implied in \cite{sparta_notional,sparta_riskassessment,bailey2021cybersecurity} but the algorithm is given here for the first time. 
Line 1 {\em subjectively} determines the SPARTA techniques that can incur risk to the space infrastructure/system in question.
Lines 2-9 assess each applicable technique, where:
lines 3-5 {\em subjectively} generate the tailored risk score associated with each applicable SPARTA technique, by determining the impact and likelihood of each technique according to the specific environment/conditions of the space infrastructure/system and mapping it to the 5$\times$5 risk matrix;
and, lines 6-9 determine if a SPARTA technique is tolerable; if not, {\em subjectively} select countermeasures and security controls to mitigate the intolerable SPARTA techniques.

\section{Characterizing NRS}
\label{sec:characterizing-v1.0}


\subsection{Case Study Based on a Real-World Attack}

The attack scenario is obtained from the authors of \cite{ear2023characterizing}.
In 2008, a cyber attack successfully compromised a NASA-managed remote sensing space system, which runs across the ground, link, and space segments but the attack objective was
the Terra satellite \cite{fritz2013satellite}---part of an earth observation system 
program for terrestrial climate research in low-earth orbit 
\cite{terra}. 
 The attack exploited an undisclosed vulnerability in the
commercially operated ground station, allowing the attacker to command the Terra satellite. 
A report \cite{Liebowitz2011Hackers} indicated that the entry point of the attack was at the Svalbard Satellite Station in Spitsbergen, Norway through an Internet connection regularly used to transfer data. However, the operator of the ground station, Kongsberg Satellite Services, found no indicators of compromise at the ground station \cite{Wolf2011China}.
The effects of the cyber attack were observed twice: satellite operators experienced interference for more than 2 minutes in June, and then more than 9 minutes in October \cite{usReport2011}. 
This allowed investigators to conclude that the attacker gained control over
the satellite, but did not conduct further exploitation \cite{Kurtin2011Satellite}. 

The attack can be described by two attack flows: one is to gain control of the satellite bus system via the network access module of the remote terminal component of the ground segment, and the other is
to conduct a denial-of-service attack against the reception module of the ground station. In what follows we show if the following space cyber risk management method was conducted when designing Terra or assigning the mission to it,
the attack (i.e., the two attack flows) 
could have been thwarted.

Figure \ref{fig:terra-flows} illustrates the mission control and data flows of Terra, the two attack flows derived from the attack, one hypothesized attack flow, and the attack techniques
used by the attacks, overlaid on top of the space system model. 
(i) One sets of mission control/data flows, which are highlighted with light blue and blue colors, depict the modules and connections required for Terra to accomplish its mission. The flight mission control flow is required to fly and position the satellite appropriately; and, the mission data flow is required to acquire and transmit sensor data to the users of the satellite. (ii) Three sets of attack flows, which are highlighted with dark red, orange, and red colors, depict the modules and connections required to successfully execute three possible attacks against Terra. 
The attack flow against the payload is not derived from the attack),
but considered for illustrating other threats can be relevant (i.e., attack modifying sensor data at the payload processing module of the data processing center component).
The case study proceeds according to Algorithm~\ref{alg:NRS} as follows.

 \vspace{-1.2em}
\begin{figure}[!htbp]
\centering{\includegraphics[width=\columnwidth]{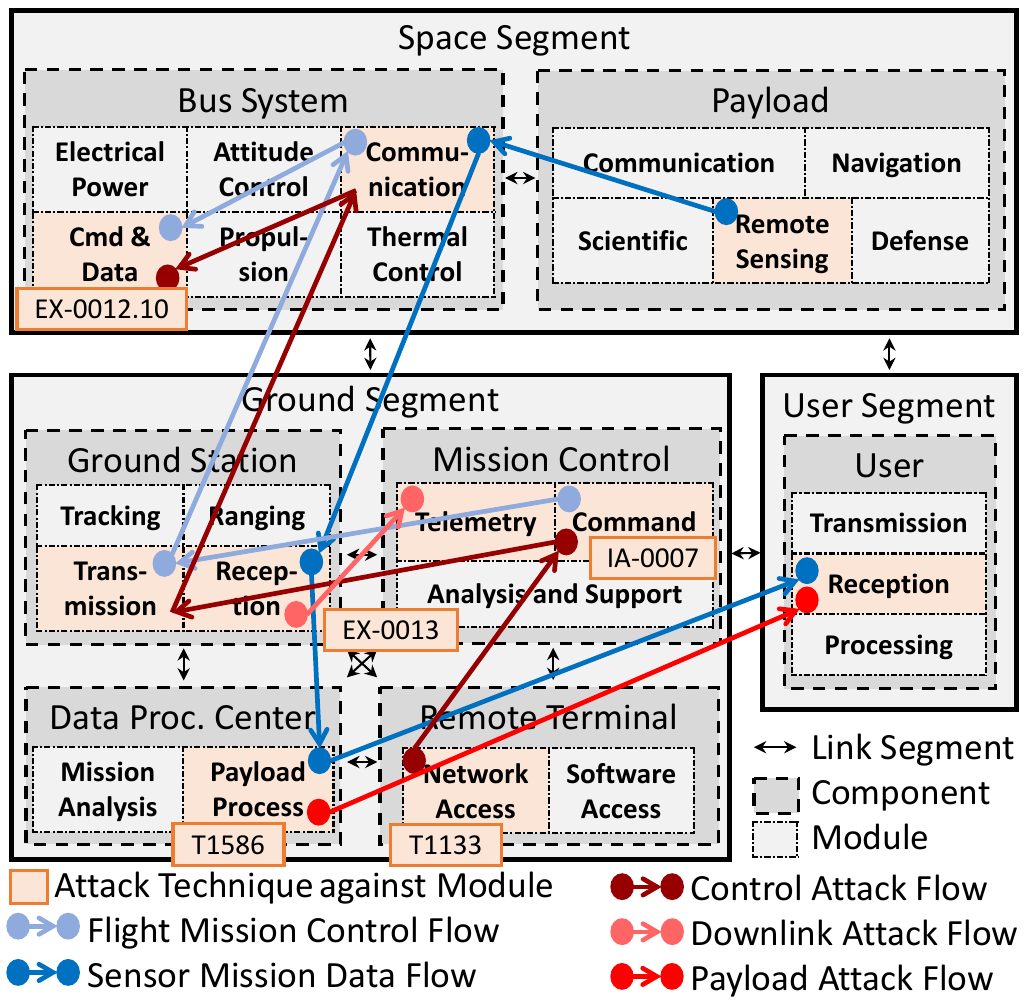}}
\vspace{-2em}
\caption{System model representing Terra overlaid with mission control/data flows, 
attack flows, and attack techniques.
}
\label{fig:terra-flows}
  \vspace{-.8em}
\end{figure}

Corresponding to Line 1 of Algorithm~\ref{alg:NRS}, suppose the analyst subjectively applies historical cyber threat intelligence (CTI), if applicable, to identify three possible attack objectives: (i) gaining control of the {\em bus system} component via the attack chain of SPARTA techniques {\em IA-0007} and {\em EX-0012.10} and ATT\&CK technique {\em T1133}; (ii) conducting a downlink disruption against the ground station via the attack chain of a single SPARTA technique {\em EX-0013}; and, (iii) compromising the integrity of the payload sensor data at the {\em user} component in the user segment,  via the attack chain of a single ATT\&CK technique {\em T1586}. Note that these attack chains and techniques are subjectively determined.

Corresponding to Lines 2-5, the analyst needs to determine the risk scores
of attacks mentioned above. Figure~\ref{fig:terra-flows} shows how 
the attack techniques of the attack chains are mapped to the modules that they directly impact; i.e., {\em EX-0012.10}, {\em EX-0013}, {\em IA-0007}, {\em T1586}, and {\em T1133} impacting the {\em command and data}, {\em reception} (at the ground station), {\em command}, {\em payload processing}, and {\em network access} modules, respectively. Then, the analyst needs to determine the criticality of these affected modules.
Suppose the analyst subjectively assesses the criticality of all modules required to accomplish the Terra mission as {\em high}, meaning that the modules affected by the attacks are all marked as {\em high} to the mission. This allows the analyst to attain base scores from NRS for each technique (Line 3). Suppose the analyst determines the attacker as Tier VI
(i.e., highly capable nation state actor) as the adversary capability (e.g., based on the nature of the mission and/or CTI). To tailor the risk scores (Line 4) correspondingly, the analyst increases (i) the {\em likelihood}, say increasing {\em IA-0007} from a likelihood of 4 to 5; and,
(ii) the {\em impact} by assigning {\em T1133} an {\em impact} of 5, because 
the network access module permits attackers to easily pivot across the network and T1133 is not assigned a base score in NRS.
The 5$\times$5 matrix is used to determine risk scores from the likelihood and impact values for each technique (Line 5). Figure~\ref{fig:5x5terra} depicts the 5$\times$5 matrix where four SPARTA and ATT\&CK techniques ({\em EX-0013}, {\em IA-0007}, {\em EX-0012.10}, and {\em T1133}) are in the {\em high} risk region, and one ATT\&CK technique ({\em T1586}) is in the {\em medium} risk region.

\vspace{-1em}
\begin{figure}[!htbp]
\centering{\includegraphics[width=.8\columnwidth]{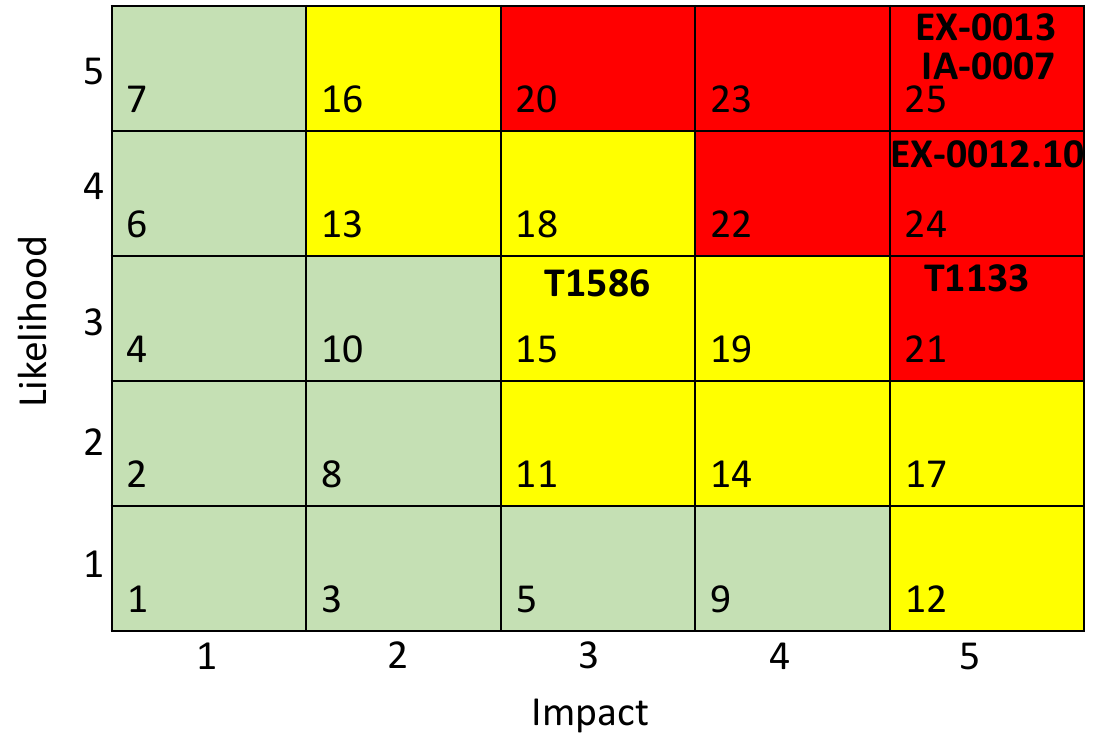}}
\vspace{-1em}
\caption{5$\times$5 risk matrix for the Terra space system.}
\label{fig:5x5terra}
\end{figure}
 \vspace{-.8em}

Corresponding to Lines 6-9, suppose the analyst decides that {\em medium} level of risk is tolerable but {\em high} level of risk is not.  Figure~\ref{fig:5x5terra} shows that the analyst must employ countermeasures to mitigate SPARTA techniques {\em EX-0013}, {\em IA-0007}, and {\em EX-0012.10}, and ATT\&CK technique {\em T1133}, but not {\em T1586} (corresponding to Line 6). For example, {\em EX-0012.10} is mapped to six countermeasures, of which countermeasure {\em CM0039} is selected to assure the least privilege principle is enforced in the command and data module (Line 7). There are 27 NIST security controls that can be used to fulfill this countermeasure. Suppose the analyst selects the following security control (Line 8) because it would be sufficient to fulfill the countermeasure: {\em CM-7}, which specifies the limitation of functions, ports, protocols, software, and services to only mission essential capabilities. Consequently, the command and data module of Terra (i.e., the bus system component of the space segment) should be able to resist attacks exploiting technique {\em EX-0012.10}.

In summary, NRS-based cyber risk management would have allowed an analyst to identify four techniques (i.e., {\em EX-0013}, {\em IA-0007}, {\em EX-0012.10}, and {\em T1133}) that must be mitigated with appropriate countermeasures and security controls.
Given that these four attack
techniques correspond to the control/downlink attack flows that are drawn from the real-world attack, the attack could have been prevented. Note that the countermeasures and security controls do not thwart 
the third attack 
mentioned above (i.e., the attack against the payload) because it was deemed tolerable risk. 

\subsection{Characteristics of NRS}

\subsubsection{Strengths of NRS}
First, it provides a much-needed starting point for practitioners to incorporate principled cyber risk management into space system design, development, operation, and maintenance. This is important because space system engineers tend to superficially assess cyber risks as low \cite{knez2016lessons}. 
As shown above, NRS can be used to identify countermeasures and security controls to mitigate risks that cannot be tolerated.

Second, it provides a baseline for space cyber risk assessment.
The large number of SPARTA techniques means a large workforce of DCO experts is required to understand and assess them all. The baseline provided by NRS decreases this workforce burden by identifying the SPARTA techniques most critical to begin cyber risk assessment (i.e., SPARTA techniques with {\em high} score values). For example, we were able to immediately identify SPARTA techniques {\em EX-0012.10} and {\em EX-0013} as priority because of their base scores given by NRS.

\subsubsection{Weaknesses of NRS}


First, it does not give justification on how The Aerospace Cooperation experts 
assessed the likelihood that a SPARTA technique would succeed against a space infrastructure or system, 
the impact incurred by the successful use of a SPARTA technique, nor 
the criticality of components or modules.
Consequently, it provides no precedent for practitioners to tailor the likelihood, impact, and criticality to their space system in question. Practitioners are left to subjectively apply their own domain expertise to tailor the base scores.

Second, significant domain expertise from practitioners is required to employ NRS to assess and mitigate cyber risk because it does not provide specific guidance for threat modeling, risk threshold establishment, nor countermeasures and security controls selection. For example, in the preceding discussion SPARTA technique {\em EX-0013}  
was subjectively selected, as well as the specific countermeasures and security controls.
%

Third, risk is not tied to attack chains or attack flows as NRS only considers risk in terms of individual SPARTA techniques but not specifying the risk associated with attackers that successfully progress from an attack entry point to a destination point via an attack flow. As a result, practitioners must subjectively aggregate the risk scores of the individual SPARTA techniques to derive the risk of an attack.  Hence, they must subjectively relate individual SPARTA techniques to components and modules 
in order to consider a bigger picture of attacks.



\subsubsection{Applicability of NRS}
NRS is applicable to space infrastructures/systems that are subject to cyber attacks, throughout their systems' lifecycle. However, this broad applicability is based on the premises that are related to the {\em subjective} capabilities indicated in Algorithm \ref{alg:NRS}. 
First, an analyst knows how to identify the attack techniques that are applicable to a space system in question. This may be difficult because it requires in-depth knowledge of adversary capabilities (across hundreds of SPARTA and ATT\&CK techniques) and how they relate to the modules and components of the space system. 

Second, an analyst knows how to adjust the base risk score of each technique and how to select appropriate countermeasures. These may be difficult because aerospace experts are required to understand the technical operation of the various modules of the space system while cybersecurity experts are required to understand the impact of cyber attacks against, and countermeasures to protect, those systems. 

Third, an analyst knows how to select appropriate security controls to instantiate countermeasures. This may be difficult because the analyst must understand how security controls originally developed for IT networks can apply to a space system.


\ignore{
\subsection{Space Cyber Attack Scenario 2}

\subsubsection{The Attack}

\subsubsection{Using the Notional Risk Scores to Analyze the Risks}

\noindent{\bf Strengths of the Notional Risk Scores}.

\noindent{\bf Weaknesses of the Notional Risk Scores}.
}




\section{Desired Properties of Space Cyber Notional Risk Scores}
\label{sec:desired-notional-risk-score}

The characterization of
the current NRS
prompts us to propose the desired properties that should be possessed by an ideal NRS,
specified in three categories: usability, usefulness, and robustness. 

\subsection{Properties Related to Usability}

We define three usability properties: {\em motivation}, {\em simplicity}, and {\em automation}. 

\begin{property}[motivation]
\label{property:clear-purpose}
An ideal NRS should have clearly defined customers or audiences.
\end{property}

Property \ref{property:clear-purpose} is important because different customers may be concerned with different perspectives of space infrastructures and systems. For example, space infrastructure developers have different considerations than space component developers, and likewise
than space module developers. Moreover, infrastructure/component/module developers have different considerations than space infrastructure/component/module operators. The purpose of an NRS should be clearly documented with example use cases, as detailed as possible. 

\begin{property}[simplicity]
\label{property:simplicity}
The simpler an NRS, the better.
\end{property}

Property \ref{property:simplicity} is important because a simple NRS would have a better chance for adoption by practitioners.
To achieve this, an NRS should be intuitive. This includes the task of preparing the input to an NRS, such as making it easy for them to (i) understand why they need to prepare certain kinds of input in order to apply an NRS, and (ii) interpret the output of an NRS and employ the guidance indicated by the output. Moreover, an NRS should make any involved technical details transparent to them, while making it easy to explain that the process for turning the input into the output is technically sound.

\begin{property}[automation]\label{property:automation}
An NRS should provide automated tools to support each  customer or audience.
\end{property}

Property \ref{property:automation} is important because manual processes are subjective to a large extent, meaning that different analysts would lead to inconsistent, if not contradicting, results. Automated analyses can reduce, if not completely avoid such subjectivity. Moreover, automation is necessary when dealing with a significant amount of data.
This is certainly the case as demonstrated in prior studies, such as \cite{knez2016lessons,ear2023characterizing}. In particular, real-world data collection processes often lack a lot of important details, forcing researchers to extrapolate the possible attack chains/flows,
and causing an explosion in the number of possible attack chains/flows
\cite{ear2023characterizing}.

\subsection{Properties Related to Usefulness}

We specify three such properties: {\em mission-centric}, {\em backward-compatibility}, and {\em broad-applicability}.

\begin{property}[mission-centric]
\label{property:tie-to-mission}
An NRS should allow one to analyze the risks associated with missions.
\end{property}

Property \ref{property:tie-to-mission} is important because space infrastructures are complex, meaning that there may always be some components that are compromised at any point in time. However, the compromises at a given point in time may only affect some missions but not others. Therefore, knowing which missions are affected by cyber attacks will direct the defenders' focus during a particular period of time, which would be essential and much more practical than aiming to secure the entire space infrastructure at every point in time. That is, we advocate {\em mission-centric} over {\em infrastructure-oriented} because the latter is too broad to be practical.

\begin{property}[backward-compatibility]
\label{property:backward-compatibility}
An NRS should accommodate, or co-operate, with existing government/industry standards and best practices.
\end{property}

Property \ref{property:backward-compatibility} is important because government standards, such as those by NIST \cite{NISTCSF}, and industry standards, such as ATT\&CK \cite{strom2018mitre}, SPARTA \cite{sparta_homepage} mentioned above, and the NASA cybersecurity best practices \cite{NASA-SS-BPG}, may be employed by the space sector. Therefore, we need to reuse their investment as much as possible. Nevertheless, we should note that the existing government/industry standards may contain weaknesses or even flaws by which an NRS should avoid being misled.

\begin{property}[broad-applicability]\label{property:broaderApplicability}
An NRS should be applicable when dealing with unknown vulnerabilities, rather than known vulnerabilities only.
\end{property}

Property \ref{property:broaderApplicability} is important because there are vulnerabilities that are not known to the defenders. Knez et al. \cite{knez2016lessons}, based on a notional scenario study, concluded that expanding threat models to include latent vulnerabilities led to a more effective space cyber risk analysis.

\subsection{Properties Related to Robustness}
We define five such properties: {\em objectivity}, {\em rigorousness}, {\em validity}, {\em uncertainty quantification}, and {\em dynamics}.

\begin{property}[objectivity]\label{property:objective}
An ideal NRS should be as objective as possible.     
\end{property}

Property \ref{property:objective} is important because it reduces unnecessary subjectivity, which could be error-prone. 
However, as shown in Algorithm \ref{alg:NRS}, the application of the current NRS is fully subjective, including how to tailor risk scores to different attacks. Subjectivity can be problematic because for example, 
space engineers lacked cybersecurity expertise and often estimated cyber risk as low \cite{knez2016lessons}.

\begin{property}[rigorousness]
\label{property:rigorousness} An ideal NRS should be based on rigorous design and analysis.
\end{property}

Property \ref{property:rigorousness} is important because the design and analysis of an ideal NRS should be adequately rigorous. This is so because rigorousness allows the clear description of assumptions made and under what conditions an NRS may not be used, while minimizing the chance of misinterpretation and ambiguity.  

\begin{property}[validity]\label{property:validity}
An NRS should be valid in terms of its models, algorithms, methods and parameters.
\end{property}

Property \ref{property:validity} is important because an unvalidated model or method and arbitrarily assigned parameter values could incur misleading results. It would be ideal that all the models, algorithms, methods, and parameters are validated through real-world datasets or real-world attack-defense experiments. However, even so, one caveat is that the data reflecting past attacks may not be reflective of the current attacks, which is believed to be an inherent limitation of data-driven characterizations. One approach to mitigating the issue is to leverage predictions of the evolution of attack capabilities and consider a distribution of predicted attack capabilities. This necessarily leads to the notion of {\em uncertainty quantification}. 

\begin{property}[uncertainty quantification]
\label{property:uncertainty-quantification}
An NRS should provide quantitative measurement associated with its output (e.g., recommendations).
\end{property}

Property \ref{property:uncertainty-quantification} is important because the output produced by an NRS may have inherent uncertainty associated with (i) its model, algorithm, and/or method as a result of approximating the real-world cyber risks; and, (ii) the parameters that are used in its model, algorithm, and/or method because there might be errors in measuring the values of parameters.  
The uncertainty associated with the output of an NRS should be taken into consideration, for example, when a human decision-maker leverages the NRS as an assistant. 

\begin{property}[dynamics]
\label{property:dynamics}
An NRS should be able to deal with the evolution of risks.
\end{property}

Property \ref{property:dynamics} is important because the space cybersecurity landscape, which includes both attack capabilities and defense capabilities, evolves with time, meaning that the risk resulting from the interactions between these dynamic attack capabilities and dynamic defense capabilities should evolve with time. 
The importance of explicitly considering dynamic risks has been discussed in a broader context than space cybersecurity (e.g., \cite{XuMTD2020}).

\section{A Formalism for Designing Future NRS}
\label{sec:notional-risk-score-v2}

Now we propose a formalism 
addressing Properties \ref{property:tie-to-mission} (mission-centric) and \ref{property:rigorousness} (rigorousness).
It is centered at the notion of {\em missions} through the lenses of {\em mission control flows} and {\em mission data flows} informally described above, meaning that we will use these flows to define the notion of mission. Informally, a mission is a specific task that needs to be accomplished,
with or without humans in the loop (e.g., humans are part of the mission) or on the loop (e.g., humans monitor the mission and make adjustment as needed),
where the task may be described at a high level that is suitable for human comprehension.  



\begin{definition}[mission control flow]
\label{def:mcf}
A mission control flow  
describes how
commands (or instructions) are issued between space infrastructure {\em units} (e.g., segment, component, module) for enabling a mission. 
Formally, a mission control flow can be represented in the language of graph theory as a {\em directed line graph}, denoted by $G_{\mcf}=(V_{\mcf},E_{\mcf})$ where each node (i.e., vertex) $v\in V_{\mcf}$ represents a space infrastructure unit and each arc $e\in E_{\mcf}$ indicates that commands are issued from one node to its immediate subsequent node in $G_{\mcf}$. 
\end{definition}

Definition \ref{def:mcf} has several features. First, it goes much beyond the current NRS, which can only be implicitly tied to the notion of missions at best. 
Second, it can describe the functional requirements associated with a  mission, as part of ``know yourself'' because it serves as a building-block for modeling cyber risks associated with a mission. Third, it is granularity- or resolution-neutral, meaning that the term {\em unit} in Definition \ref{def:mcf} can be equally applied to describe mission control flows at the {\em segment} level (e.g., a mission requires the ground segment to send commands to the space segment for execution), at the {\em component} level
(e.g., a mission requires the {\em mission control} component of the ground segment to send commands to the {\em bus system} component of the space segment), and at the {\em module} level (e.g., a mission requires the {\em command} module in the {\em mission control} component at the ground segment to send commands to the {\em command and data} module in the {\em bus system} component at the space segment). Fourth, in principle, it can be drilled down to the traditional notion of control flow in program analysis. However, we do not advocate this lowest level of abstraction because it is not clear how it can capture the meaning of {\em mission}, which is by default at a high level of abstraction. 
Fifth, since a mission may incur many mission control flows, it naturally leads to the notion of {\em mission control flow graph}:

\begin{definition}
[mission control flow graph] 
\label{def:mcfg}
Given a set of $m$ mission control flows incurred by a mission, denoted by
$G_{\mcf}^{(i)}=(V_{\mcf}^{(i)},E_{\mcf}^{(i)})$ where $i=1,\ldots,m$, 
a mission control flow graph is denoted by $G_{\mcfg}=(V_{\mcfg},E_{\mcfg})$, where $V_{\mcfg}=\bigcup_{i=1}^m V_{\mcf}^{(i)}$ and $E_{\mcfg}=\bigcup_{i=1}^m E_{\mcf}^{(i)}$.
\end{definition}

Note that a mission control flow graph is not a straightforward combination of a set of mission control flows, for several reasons. First, two mission control flows may share one or multiple nodes and/or arcs, explaining why it is a graph in general. Second, there may be some nodes whose in-degrees are zero because they initiate commands. Third, there may be some nodes whose out-degrees are zero because they do not initiate any command to another node but simply execute commands. Fourth, in principle, a mission control flow graph would be acyclic, or Directed Acyclic Graph (DAG); otherwise, a mission may never complete because of a loop.

Similarly, we define the notion of mission data flow and mission data flow graph, while highlighting the differences from their control flow counterparts.

\begin{definition}[mission data flow] 
\label{def:mdf}
A mission data flow 
describes 
how data items are transmitted
between space infrastructure {\em units} (e.g., segment, component, module) for enabling a mission. Formally, a mission data flow can also be represented in the language of graph theory as a {\em directed line graph}, denoted by $G_{\mdf}=(V_{\mdf},E_{\mdf})$ where each node (i.e., vertex) $v\in V_{\mdf}$ represents a space infrastructure unit and each arc $e\in E_{\mdf}$ indicates that data items are sent from one node to its immediate subsequent node in  $G_{\mdf}$. 
\end{definition}

Note that a mission data flow may have something in common with a mission control flow, possibly $V_{\mcf}\cap V_{\mdf} \neq \emptyset$ because the nodes in both sets represent space system units. For example, $(v_1,v_2)\in E_{\mcf}$ may represent that node $v_1$ commands $v_2$ to send the sensor data collected at $v_2$ to $v_1$,
meaning $(v_2,v_1)\in E_{\mdf}$. 

\begin{definition}
[mission data flow graph] 
\label{def:mcfg}
Given a set of $n$ mission data flows incurred by a mission, denoted by $G_{\mdf}^{(i)}=(V_{\mdf}^{(i)},E_{\mdf}^{(i)})$ where $i=1,\ldots,n$,
a mission data flow graph is denoted by $G_{\mdfg}=(V_{\mdfg},E_{\mdfg})$, where $V_{\mdfg}=\bigcup_{i=1}^n V_{\mdf}^{(i)}$ and $E_{\mdfg}=\bigcup_{i=1}^n E_{\mdf}^{(i)}$.
\end{definition}

Similar to mission control flow graphs, a node in a mission data flow graph may have in-degree 0 (i.e., the node only produces data) or out-degree 0 (i.e., the node only accepts and processes data).
Unlike mission control flow graphs, a mission data flow graph may be cyclic, but the data would be of different kinds. For example, it is possible $(v_1,v_2)\in E_{\mdfg}$ and $(v_2,v_1)\in E_{\mdfg}$. 

The preceding definitions pave a way to formalize the notion of mission.

\begin{definition}
[mission]\label{def:mission}
A mission with a mission control flow graph $G_{\mcfg}=(V_{\mcfg},E_{\mcfg})$ and mission data flow graph $G_{\mdfg}=(V_{\mdfg},E_{\mdfg})$ is described as $G_{\mission}=(V_{\mission},E_{\mission})$, where $V_{\mission}=V_{\mcfg} \bigcup V_{\mdfg}$ and $E_{\mission}=E_{\mcfg} \bigcup E_{\mdfg}$.
\end{definition}

\ignore{

\subsection{Use Cases of the Formalism}

{\color{red}if times does not permit, we may not include this subsection for this version}

The key idea behind our proposal is to tie the space-based cyber kill chain to a mission as follows: (i) What are the cyber kill chains that target each node in a mission graph $G_{\mission}=(V_{\mission},E_{\mission})$? (ii) What are the cyber kill chains that target each arc in a mission graph $G_{\mission}$? (iii) Given the risk associated with each node $v\in V_{\mission}$
and each arc $e\in E_{\mission}$, how should we aggregate them together with respect to the structure of $G_{\mission}$ to derive the mission-level risk, denoted by $R_{\mission}$? 

} 


\section{Conclusion and Future Directions}
\label{sec:conclusion}

Space cyber developers and operators need guidance to accomplish space cyber risk management. Along this line, NRS represents a good first step. We presented an algorithmic description of NRS and a first characterization on its strengths, weaknesses, and applicability. To guide the development of future NRS, we proposed a comprehensive set of desired properties that should be satisfied by future NRS. We also proposed a formalism to serve as a starting point towards developing future NRS with the desired properties and automated space cyber risk management tools.

There are two major research directions. One is to refine the design of the current NRS, or propose a completely new NRS. Along this line, open problems include: How can the base risk scores be made objective?
How can practitioners tailor space cyber risk scores objectively? 
How can an NRS support a broad range of audiences while maintaining adequate simplicity to remain intuitive? 
How can we design automated tools to help practitioners determine the criticality level and impact of each node in a mission graph (i.e., space infrastructure unit)?
Are there other desired properties?
What NRS can satisfy all the desired properties? 

The other research direction is to develop automated space cyber risk management tools while leveraging an NRS that satisfies most, if not all, the desired properties. Along this direction, open problems include: How 
should we extend the formalism to incorporate attack chains that target nodes and/or arcs in a mission graph and develop full-fledged space cyber risk management tools?
How should we leverage the (revised/extended) formalism to design principled methods to replace the heuristic 
5$\times$5 matrix method?
How can we automatically extract a mission graph from the mission specification?
How can the criticality and impact of space system units in a mission graph be automatically computed from space system specification, environmental conditions, CTI, and mission requirements? 

\bibliographystyle{IEEEtran}
\bibliography{references,metrics}

\end{document}